\documentclass{elsart}
\usepackage{graphicx,amsmath,amssymb,mathrsfs}
\journal{Physica A}

\begin{document}
\begin{frontmatter}
\title{Dynamical modelling of superstatistical complex systems}
\author{Erik Van der Straeten} and
\ead{e.straeten@qmul.ac.uk}
\author{Christian Beck}
\ead{c.beck@qmul.ac.uk}
\address{Queen Mary University of London, School of Mathematical Sciences,\\ Mile End Road, London E1 4NS, UK}

\begin{abstract}
We show how to construct the optimum superstatistical dynamical model for a given experimentally measured time series.
For this purpose we generalise the superstatistics concept and study a Langevin equation with a memory kernel
whose parameters fluctuate on a large time scale. It is shown how to
construct a synthetic dynamical model with the same invariant
density and correlation function as the experimental data. 
As a main example we apply
our method to velocity time series measured in
high-Reynolds number turbulent Taylor-Couette flow, 
but the method can be applied
to many other complex systems in a similar way.
\end{abstract}
\end{frontmatter}


\section{Introduction}
Many complex systems in nature exhibit a rich structure of dynamics,
described by a mixture of different stochastic processes on
various time scales. It has become common to call such dynamical
processes with time scale
separation {\em superstatistical} \cite{super_or,turb_su01, eco_su01,
turb_su02,touchette, prl07, jizba, anteneodo}. A fascinating feature
is that many highly nonlinear equations, for example the Navier-Stokes
equations, do effectively produce stochastic processes that can be
very well approximated by linear stochastic differential equations
with fluctuating parameters on a large time scale,
i.e., superstatistical processes. Examples where superstatistical
techniques have proved to be a powerful tool are
Lagrangian trajectories
of tracer particles in fully developed turbulence \cite{turb_su01,turb_su02},
share price dynamics with volatility fluctuations \cite{jizba, anteneodo,
eco_su01}, 
atmospheric and geophysical processes \cite{porpo},
random matrix theory \cite{abul} and random networks \cite{abe}. 
The basic idea
underlying most of these models is that locally the system is described by a
simple Gaussian stochastic process, and only on a large scale
the mixture of various Gaussian processes with fluctuating variance leads
to more complex behavior.

So far the superstatistical model building has mainly concentrated on
reproducing the correct stationary density of the nonlinear system under
consideration. However, it is certainly important to incorporate
dynamical information from the set of all higher-order correlation
functions. In fact, experimentalists are often able
to measure two relevant quantities, the stationary probability
distribution of a fluctuating time series produced by the
complex system under consideration, and the two-point correlation function
for relatively small time scale differences.
The purpose of this article is to describe how to construct the optimum
superstatistical model for a given
complex system taking these two experimentally measured 
quantities as input.
As a main example, we will construct a
suitable superstatistical model for measured velocity time series
in turbulent Taylor-Couette flow, which is in
excellent agreement with the experimental data.
Our techniques presented here are quite generally
applicable to large classes of complex systems where the local
dynamics is well approximated by a generalised Langevin equation with
memory kernel.

\section{Standard Langevin equation}
We start from a simple example that will be
subsequently
generalised. Consider locally a linear stochastic differential
equation (without memory kernel) of the form
\begin{equation}\label{lan_stan}
\frac{d}{dt}v(t)=-\gamma v(t)+\sqrt{\frac{2\gamma}\beta}g(t),
\end{equation}
where $\gamma$ and $\beta$ are constants and $g(t)$ is Gaussian
white noise. This generates the Ornstein-Uhlenbeck process, a Gaussian
Markov process, with stationary distribution $p(v)$ and 
correlation function $c_v(\Delta)$:
\begin{eqnarray}
p(v)&=&\sqrt{\frac\beta{2\pi}}e^{-\frac12\beta v^2}
\cr 
c_v(\Delta)&=&\lim_{t\rightarrow\infty}\langle v(t)v(t+\Delta)
\rangle=\frac1\beta e^{-\gamma\Delta}.
\end{eqnarray}
$\langle.\rangle$ denotes an ensemble average. In the superstatistical approach, one regards 
the parameter $\beta$ as a random
variable that fluctuates on a large time scale.
If this random variable has the probability distribution
$f(\beta)$, then the stationary (marginal) distribution of the complex
system under consideration is given by
\begin{equation}\label{here}
P(v)=\int_0^\infty d\beta f(\beta)\sqrt{\frac\beta{2\pi}}e^{-\frac12\beta v^2}.
\end{equation}
All kinds of distributions $P(v)$ can be generated in this way,
for example power-law decays in $v$ or stretched exponentials,
depending on the properties of $f(\beta)$ \cite{touchette}.
In fact, an experimentally measured $P(v)$ can be used to construct
$f(\beta)$ from eq.~(\ref{here}).
The superstatistical correlation function is given by
\begin{equation}\label{fbcv}
C_v(\Delta)=\int_0^\infty d\beta f(\beta)c_v(\Delta)=e^{-\gamma\Delta}\int_0^\infty d\beta f(\beta)\frac1\beta\propto c_v(\Delta).
\end{equation}
Clearly, an experimentalist may ask whether such a simple
model is appropriate for a given experimental data set.
In most cases it will not, because the measured 
decay of the correlation function is not simply given by
$\exp(-\gamma\Delta)$ as is the case for the Ornstein-Uhlenbeck process.
In fact, experimental data often show local correlation functions that
are well described by a sum or difference of two exponentials,
or sometimes there is oscillating behavior. Our aim
in the following is thus to extend the local dynamics
in such a way that this more refined local dynamics,
reproducing the correct (measured) correlation function, can be
incorporated into the superstatistical modelling approach.

\section{Generalised Langevin equation}
A more general starting point is to consider
locally a set of coupled linear stochastic differential equations
\begin{equation}\label{set}
\frac{d}{dt}v_j(t)=-\sum_k\Theta_{j,k}v_k(t)+\Xi_jg_j(t).
\end{equation}
The elements of the matrix $\Theta$ and the vector $\Xi$ are constants. The noise
terms are denoted by $g_j(t)$. The idea of the theory of Mori-Zwanzig \cite{zwan,mori} 
is that some of the variables $v_j(t)$ are irrelevant,
or inaccessible to the experimentalist.
By eliminating these variables out of the equations of motion, 
one ends up with a set of {\it non-Markovian} equations in the 
relevant variables only. In this article we study in detail  
a 2-dimensional version of this problem, namely 
\begin{eqnarray}\label{exlan1} 
\frac{d}{dt}v(t)&=&-\gamma v(t)+V(t)+\alpha g(t)
\\\label{exlan2} 
\frac{d}{dt}V(t)&=&-\eta V(t) +\sigma v(t).
\end{eqnarray}
Here, $\eta$ and $\gamma$ are positive constants. The constants $\alpha$ and 
$\sigma$ can have either sign. $g(t)$ denotes Gaussian white noise, i.e.,
$\langle g(t)\rangle=0$ and 
$\langle g(t)g(t')\rangle=\delta(t-t')$. For the above dynamics, 
the two eigenvalues of the matrix $\Theta$ are $A\pm B$, with
\begin{equation}\label{AB}
2A=\eta+\gamma\quad\textrm{and}\quad2B=\sqrt{(\eta-\gamma)^2+4\sigma}.
\end{equation}
It is easy to see that the real parts of these eigenvalues are always 
positive provided $\eta\gamma-\sigma>0$. In the remaining part of this article, we assume that this inequality holds. This insures that the system will 
locally reach a stationary state \cite{zwan}.

Our physical interpretation is as follows: 
The velocity variable $v(t)$ decays with relaxation constant 
$\gamma$ and is subjected to two forces: A rapidly fluctuating force $g(t)$ that drives the dynamics and another force
$V(t)$ that decays with relaxation constant $\eta$. This force $V(t)$,
however, is not independent of the velocity $v(t)$. Rather, the
change of $V(t)$ is proportional to the velocity $v(t)$,
with proportionality constant $\sigma$. The above linear
coupling scheme is one of the simplest, but
non-trivial, examples of coupled local dynamics
between pairs of variables. Now assume that an experimentalist can only measure one of these two variables. We chose $v(t)$ to be this variable and proceed by eliminating $V(t)$ out of the set of equations~(\ref{exlan1},\ref{exlan2}). In this way one obtains an equation in $v(t)$ with a memory kernel \cite{zwan,mori}:
\begin{equation}\label{gle_af}
\frac{d}{dt}v(t)=-\gamma v(t)+\sigma\int_0^{t}dt'e^{-\eta(t-t')}v(t')+\alpha g(t).
\end{equation}
To obtain this formula, we ignored the contribution $V(0)\exp(-\eta t)$
because this term is unimportant for long times. Expression (\ref{gle_af}) contains an exponential memory kernel and is known in the literature as the generalised Langevin equation \cite{zwan,mori}. Usually, one assumes that the time dependence of the memory kernel and the correlation function of the stochastic term are equal. This equality is called the fluctuation-dissipation relation of the second kind \cite{Kubo_fluc} and the stochastic term $g(t)$ is then interpreted as internal noise. However, for the example studied in the present paper, the aforementioned time dependencies are not equal. Then, the stochastic term $g(t)$ is interpreted as external noise. Such noise terms have also been studied recently in the context of stochastic resonance \cite{harm_ocs} and anomalous diffusion \cite{ano_diff}.

Two well-known special cases are included in our local
model. Taking the limit $\eta\rightarrow\infty$ in~(\ref{gle_af}) results 
in the standard Langevin equation~(\ref{lan_stan}) without
memory kernel. Equivalently,
this case also arises by putting $\sigma =0$, i.e.\ no coupling
between velocity and force. Moreover,
expressions~(\ref{exlan1},\ref{exlan2}) reduce to the equations of motion 
of a Brownian harmonic oscillator for $\eta=0$ and $\sigma<0$. 
The variables $v(t)$ and $-V(t)$ are then interpreted as the 
velocity and the position of the oscillator, respectively.

\section{Local model: stationary distribution}
Before implementing superstatistics, we first have to solve the above local model. Let us start with calculating the local stationary distribution  $p(v)$ of the relevant variable $v(t)$. The Fokker-Planck equation \cite{kamp} for the joint probability distribution $p(v,V,t)$ is
\begin{equation}
0=\bigg[\frac{\partial}{\partial t}+\frac{\partial}{\partial v}(V-\gamma v)+\frac{\partial}{\partial V}(\sigma v-\eta V)-\frac{\alpha^2}2\frac{\partial^2}{\partial v^2}\bigg]p(v,V,t).
\end{equation}
The stationary distribution $p(v,V)$ is obtained by setting
 $\partial p(v,V,t)/\partial t=0$ and is given by
\begin{equation}
p(v,V)\sim\exp\left[-\frac{\gamma+\eta}{\sigma^2\alpha^2}\left[\left(\sigma v-\eta V\right)^2+\left(\eta\gamma-\sigma\right)V^2\right]\right].
\end{equation}
For the marginal distribution $p(v)$ we integrate out the dependence on $V$ 
\begin{equation}\label{locgau}
p(v)=\int_{-\infty}^{+\infty}dVp(v,V)=\sqrt{\frac\beta{2\pi}}e^{-\frac12\beta v^2},
\end{equation}
where the parameter $\beta$ is given by
\begin{equation}\label{kappa_stat}
\beta=\frac{2(\gamma+\eta)(\eta\gamma-\sigma)}{\alpha^2(\eta\gamma-\sigma+\eta^2)}. 
\end{equation}
We see, as required for superstatistics, that
the local stationary distribution
is a Gaussian distribution.

\section{Local model: correlation function}
\begin{figure}
\begin{center}
\includegraphics[width=0.5\textwidth]{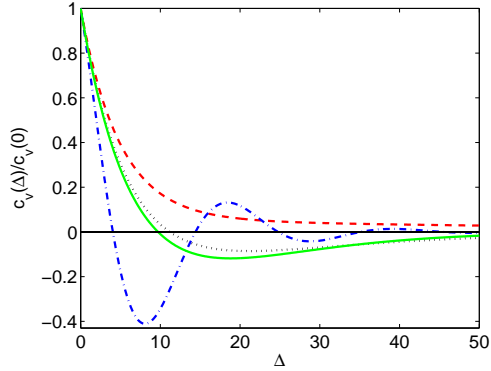}
\caption{\label{fig}(Colour online) Plot of the correlation function 
for different values of $\sigma$. The solid line is a representation of 
expression~(\ref{Bzero}), i.e.\ the case $B=0$.
The dashed-dotted curve shows an oscillatory 
behavior~(\ref{osc}). The dashed and dotted curve are respectively a sum and difference of two exponential functions~(\ref{sum_diff}).}
\end{center}
\end{figure}
To proceed further into a dynamical setting, we have to
calculate a formal solution for $v(t)$.
Let us denote the Laplace transform $\hat f(w)$ of a function $f(t)$ by
\begin{equation}
\hat f(w)=\int_0^\infty dt e^{-wt}f(t).
\end{equation}
From eqs.~(\ref{exlan1},\ref{exlan2}) one obtains $\hat v(w)$ as
\begin{equation}\label{fourvw}
\hat v(w)=\hat\varrho(w)\left[\frac{V(0)}{w+\eta}+v(0)+\alpha\hat g(w)\right],
\end{equation}
where the function $\hat\varrho(w)$ is
\begin{equation}\label{varp}
\hat\varrho(w)=(w+\eta)(w+A+B)^{-1}(w+A-B)^{-1}
\end{equation}
(see~(\ref{AB}) for the definition of $A$ and $B$). By calculating the inverse Laplace transform of~(\ref{fourvw}), 
one obtains
\begin{equation}
v(t)=v(0)\varrho(t)+V(0)\int_0^tdt'\varrho(t-t')e^{-\eta t'}+\alpha\int_0^tdt'\varrho(t-t')g(t').
\end{equation}
The stationary correlation function $c_v(\Delta)$ is obtained by 
evaluating the ensemble average of the product of $v(t)$ and 
$v(t+\Delta)$ for $t\rightarrow\infty$. All the terms involving the 
initial conditions $v(0)$ and $V(0)$ that contribute to $c_v(\Delta)$ are 
vanishing for $t\rightarrow\infty$ and in this limit one ends up with the
following expression for the correlation function:
\begin{equation}\label{defcorr}
c_v(\Delta)=\alpha^2\int_0^\infty d\tau\varrho(\tau)\varrho(\tau+\Delta).
\end{equation}
In order to obtain an explicit expression for $c_v(\Delta)$, 
we have to calculate the inverse Laplace transform of 
$\hat\varrho(w)$ given by (\ref{varp}). The result of this transformation depends on whether 
$B=0$ or $B\neq0$. However it is easy to prove that the following 
inequalities hold: $\mathscr{R}(A\pm B)>0$ given $\eta>0$, $\gamma>0$ and
 $\eta\gamma-\sigma>0$. This means that the Laplace transform is well-defined 
provided that the latter inequalities are satisfied. Under the assumption that $B=0$, one 
obtains the following expression for the function $\varrho(t)$
\begin{equation}
\varrho(t)=[1+(\eta-A)t]e^{-At},
\end{equation}
while for the case $B\neq0$ the result is 
\begin{equation}
\varrho(t)=\frac1{2B}(-\eta+A+B)e^{-(A+B)t}-\frac1{2B}(-\eta+A-B)e^{-(A-B)t}.
\end{equation}
We proceed by evaluating~(\ref{defcorr}) to calculate an explicit expression for the correlation function. Under the assumption that $B=0$, one obtains
\begin{equation}\label{Bzero}
\frac{c_v(\Delta)}{c_v(0)}=\frac{e^{-A\Delta}}{\eta^2+A^2}\left[\eta^2+A^2+A\left(\eta^2-A^2\right)\Delta\right].
\end{equation}
We split the case $B\neq0$ into two subcases, depending on whether $B$ is imaginary or real. In case $B\neq0$ and $\mathscr{R}(B)=0$, the result for $c_v(\Delta)$ is
\begin{equation}\label{osc}
\frac{c_v(\Delta)}{c_v(0)}=e^{-A\Delta}\left[\Gamma'\sin\left(B'\Delta\right)+\cos\left(B'\Delta\right)\right],
\end{equation}
while in case $B\neq0$ and $\mathscr{R}(B)=B$, one ends up with
\begin{equation}\label{sum_diff}
\frac{c_v(\Delta)}{c_v(0)}=\frac{e^{-A\Delta}}{1+\Gamma}\left[e^{-B\Delta}+\Gamma e^{B\Delta}\right],
\end{equation}
with $B'=B/i$ and
\begin{equation}
\Gamma'=\frac A{B'}\frac{\eta^2-A^2-B'^2}{\eta^2+A^2+B'^2},\ \ \ \ \Gamma=\frac{B+A}{B-A}\frac{(B-A)^2-\eta^2}{(B+A)^2-\eta^2}.
\end{equation}
It is easy to show that the signs of $\Gamma$ and $\sigma$ are equal. This means that the parameter $\sigma$ determines whether one ends up 
with a sum or a difference of exponentials in expression~(\ref{sum_diff}). 
As a consequence, there are $4$ qualitatively different
types of correlation functions contained in the set of stochastic 
differential equations~(\ref{exlan1},\ref{exlan2}). This is illustrated 
in Figure~\ref{fig}. The figure shows the correlation function for $4$ 
different values of the coupling parameter $\sigma$ but with constant 
values of the other parameters $\gamma=0.2$ and $\eta=0.02$. Our chosen 
example values of $\sigma$ are $-0.0081$ (solid line), $-0.1$ (dashed-dotted line), $0.001$ (dashed line) and $-0.005$ (dotted line). Experimentally measured correlation functions in turbulence
are often of strikingly similar shape, see for example Figure~2a
in \cite{boden}.

\section{Superstatistics}
Finally let us apply superstatistics to this local model
by introducing large-scale fluctuations in $\beta$. 
We assume that $\beta$, as given by eq.~(\ref{kappa_stat}), is distributed according
to some probability density $f(\beta)$ and that the changes
of $\beta$ occur on a time scale that is much larger than 
the relaxation time of the correlation function $c_v(\Delta)$.
Generally, this results in a superstatistical distribution $P(v)$
given by (\ref{here}) 
that is non-Gaussian. The time-dependence of the superstatistical 
correlation function $C_v(\Delta)$ given by (\ref{fbcv}) is equal to the 
time-dependence of $c_v(\Delta)$ as long as the $\beta$-fluctuations are 
produced by fluctuations of $\alpha$ only. 

We are now in a position to formulate the central result of this article.
Let a complex system be given that exhibits time scale separation and
superstatistical behavior (see \cite{eco_su01} for a test). Suppose we know 
$P(v)$ and $C_v (\Delta)$
from experimental measurements. How should we construct
the optimum synthetic superstatistical model that is consistent with
the measured data? Of course many nonlinear models are possible,
but if one assumes that there is a local linear coupling between
pairs of dynamical variables and that locally the stochastic process
considered exhibits
Gaussian behavior, then the model~(\ref{exlan1},\ref{exlan2}) 
with superstatistical parameter fluctuations is the most appropriate one.
The distribution $f(\beta)$ of the parameter $\beta$ can be
determined by comparing eq.~(\ref{here}) with the experimentally measured distribution $P(v)$. Then,
depending on the shape of
the measured correlation function, the values of the parameters $\gamma$, $\eta$ and $\sigma$
can be determined. 

\begin{figure}
\begin{center}
\includegraphics[width=0.5\textwidth]{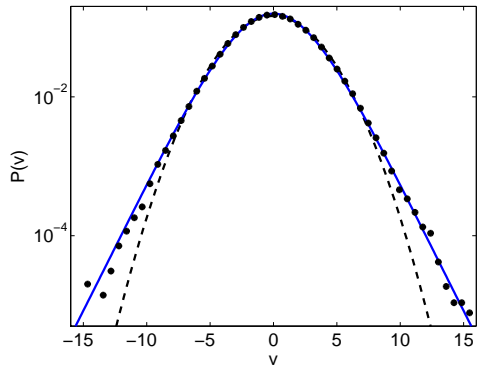}
\caption{\label{fig2}(Colour online) Plot of the measured distribution $P(v)$ ($\bullet\bullet\bullet$). The dashed line represents a Gaussian approximation $0.15\exp(-0.07v^2)$ for $P(v)$. The solid line is the superstatistical distribution~(\ref{here}), with $f(\beta)$ a lognormal distribution~(\ref{poi}).}
\end{center}
\end{figure}
\begin{figure}
\begin{center}
\includegraphics[width=0.48\textwidth]{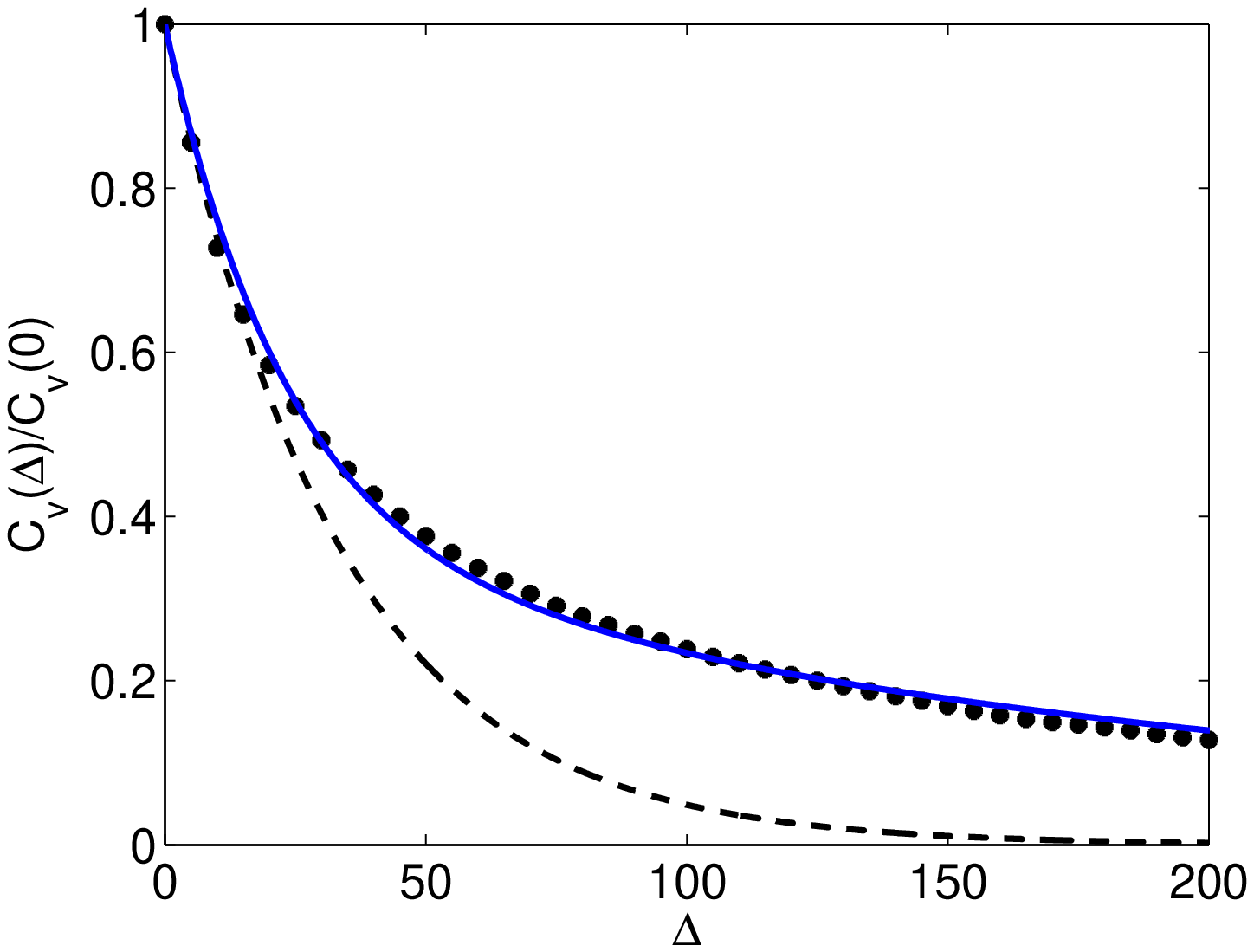}
\includegraphics[width=0.48\textwidth]{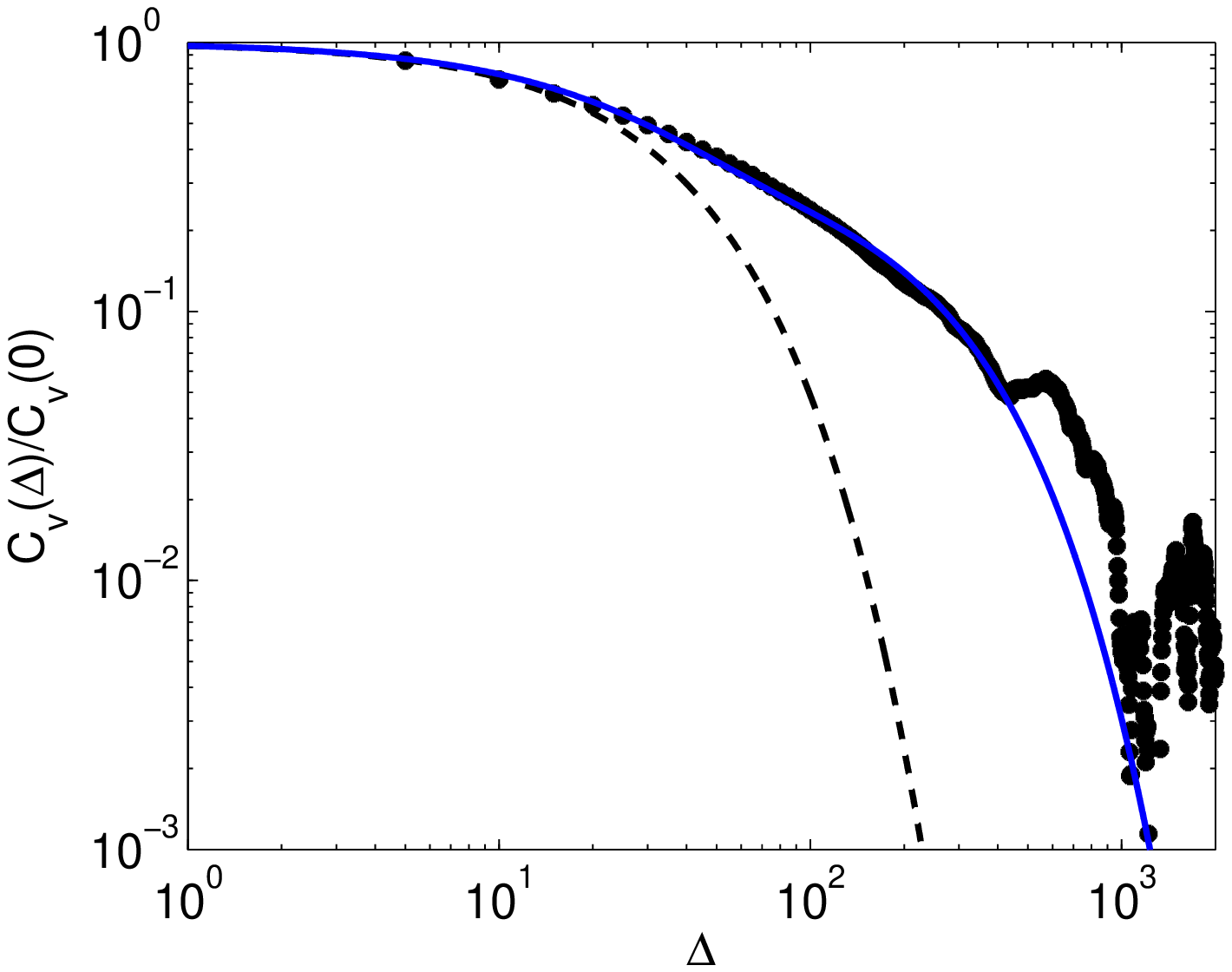}
\caption{\label{fig3}(Colour online) Plot of the measured normalised
 correlation function $C_v(\Delta)/C_v(0)$ ($\bullet\bullet\bullet$). 
The dashed and solid lines represent an approximation for 
$C_v(\Delta)/C_v(0)$ in terms of a single exponential 
$\exp(-0.030\Delta)$ and a sum of two exponentials~(\ref{sum_diff}),
respectively. Fig~3a linear plot, Fig.~3b double 
logarithmic plot.}
\end{center}
\end{figure}

To illustrate our general method,
we study as an example
time series obtained in an experiment performed by Lewis and 
Swinney \cite{exp_lew, turb_su01}. The data set contains the values of a 
single velocity component $v(t)$ as a function of time $t$ in 
turbulent Taylor-Couette flow for Reynolds number $Re=540.000$. 
The stationary probability distribution $P(v)$ exhibits 
non-Gaussian behavior, see Figure~\ref{fig2}. A suitable choice for $f(\beta)$, motivated by
the K62 theory of turbulence \cite{kolmo}, is a lognormal distribution
\begin{equation}\label{poi}
f(\beta)=\frac1{b_1\sqrt{2\pi}\beta} \exp\left(-(\ln\beta-b_2)^2/2b_1^2\right).
\end{equation}
Choosing $b_1=0.42$ and $b_2=-1.91$ one obtains an excellent fit of
the experimentally measured histogram, see
Figure~\ref{fig2}. But there are of course many
different stochastic processes generating the
same stationary distribution $P(v)$. We
can now construct a more precise synthetic model of the dynamics by
taking into account the information contained in the measured
correlation function. The experimentally measured 
correlation function $C_v(\Delta)/C_v(0)$ is shown
in Figure~\ref{fig3}a, together with
a single exponential function $\exp(-0.030\Delta)$ 
and a sum of two exponentials, see expression~(\ref{sum_diff})
 with $A=0.024;B=0.019;\Gamma=0.571$. Figure~\ref{fig3}a clearly shows that 
a single exponential function is not able to properly
fit the measured correlation function, while a sum of two exponentials 
yields an excellent fit. This means that our superstatistical memory-kernel
approach is a suitable way to model these data.
 
In this paper, we did not consider the $\beta$-process.
That is to say, we did not construct an explicit model for the
time development of the slowly varying variable $\beta$. 
However, it is clear that for very long time scales the details of this 
process will influence the correlation function $C_v(\Delta)/C_v(0)$ 
and the stationary distribution $P(v)$. 
To investigate the behaviour for larger time scales $\Delta$,
Fig.~\ref{fig3}b shows the measured and modelled correlation function
in a double-logarithmic plot, which emphasizes the long-time behaviour.
We see that our model (solid line) fits the measured data well
up to time scales $\Delta \approx 400$. Beyond that, there are
large statistical fluctuations and the behaviour is dominated by details of
the $\beta$-process. For example, an asymptotic power law decay could
be constructed by considering a $\beta$-process whose long-term
correlations decay with a power law. 
Taking the $\beta$-process into account goes beyond the  
scope of the current paper
but is an interesting topic for further research. 
Our current approach as presented in this paper gives testable
predictions for small and intermediate time scales $\Delta$.
The asymptotic $\Delta \to \infty$ behaviour is then dominated by the
chosen $\beta$-process.

Let us emphasise that obtaining a sum (and not a difference) of 
two exponentials for the correlation function~(\ref{sum_diff}) 
is non-trivial. For this it is crucial that $\sigma>0$ and as a consequence 
$\eta\neq0$ and $\gamma\neq0$. Such correlation functions are also observed in biological time series for trajectories of motile cells \cite{biol} and have been recently obtained theoretically for two-dimensional stochastic motion with uncorrelated fluctuations of the speed and the direction of the motion \cite{prop}.

\section{Discussion}
To summarise, in this article we have shown how to construct an optimum
superstatistical dynamical model that exhibits the same invariant density
and correlation function 
as an experimentally measured time series extracted
from a complex system.
The relevant universality class of complex systems
to which our approach is applicable consists of systems
where locally pairs of dynamical variables are coupled in
the simplest possible (linear) way, as described by eqs.~(\ref{exlan1},\ref{exlan2}).
This leads to a Langevin equation with memory kernel whose
parameters fluctuate on a large time scale
in a superstatistical way. As an example, we
applied our approach to experimentally measured velocity time series
in turbulent Taylor-Couette flow, obtaining excellent agreement with
the experimental data. Our approach is quite generally applicable to many 
complex systems with time scale separation, as long as 
high-precision experimental data on the stationary density
and correlation function are available.

\end{document}